\documentclass[prl,twocolumn,amsmath,amssymb,floatfix,showpacs]{revtex4}
\usepackage{graphicx}
\usepackage{bm}
\usepackage{amssymb}
\usepackage{color}
\usepackage{epsfig}
\usepackage{subfigure}

\newcommand{\be}{\begin{equation}}
\newcommand{\ee}{\end{equation}}
\begin{document}

\title {Variable range hopping in the Coulomb glass}

\author{Ariel Amir,
 Yuval Oreg and
 Yoseph Imry,
}

\affiliation{Department of Condensed Matter Physics, Weizmann
Institute of Science, Rehovot, 76100, Israel}

\begin{abstract}

We use a mean-field (Hartree-like) approach to study the conductance
of a strongly localized electron system in two dimensions. We find a
crossover between a regime where Coulomb interactions modify the
conductance significantly to a regime where they are negligible. We
show that under rather general conditions the conduction obeys a
scaling relation which we verify using numerical simulations. The
use of a Hartree self-consistent approach gives a clear physical
picture, and removes the ambiguity of the use of single-particle
tunneling density-of-states (DOS) in the calculation of the
conductance. Furthermore, the theory contains interaction-induced
correlations between the on site energy of the localized states and
distances, as well as finite temperature corrections of the DOS.

\end{abstract}


 \maketitle

 \section{Introduction}

  The unusual
temperature dependence of the conductance of strongly disordered
samples has generated a great interest in the past four decades.
These systems manifest an interplay between disorder and
interactions, which is ubiquitous in nature, and which renders the
problem difficult theoretically and rich experimentally
\cite{{efros},{efrosbook2},{Gantmakherbook},{pollakbook1}}. In these
disordered systems, conductance occurs via phonon assisted hopping
between localized states, whose size is roughly the localization
length $\xi$. The measured low-temperature conductance of various
disordered samples shows temperature dependence different than
simple activation, of the type $\sigma \sim
e^{-\left(\frac{T_0}{T}\right)^\zeta}$, with $\zeta <1$.

Mott's theory ("Mott's law") for Variable Range Hopping
(VRH)~\cite{mott}, which ignores the Coulomb interactions, gives
$\zeta=\frac{1}{d+1}$ and $T_0$ equal
 to the non-interacting mean level spacing in a localization volume $\delta_\xi=1/\nu_0 \xi^d$; with $d=1,2$ or 3 the dimension of the system and $\nu_0$ the density-of-states.

  As first pointed out by Pollak~\cite{pollak_}, interactions
affect the single-particle tunneling density-of-states (DOS), and
thus may possibly influence the conductance. Efros and Shklovskii
(ES)~\cite{efros2} obtained an analytic bound for the so-called
Coulomb gap in the DOS, which was later observed experimentally
\cite{{coulomb_gap_experimental},{butko}}. Using the
interaction-modified DOS to find the conductance (treating the
problem as essentially non-interacting from this point) one obtains
("ES's law") $\zeta=\frac{1}{2}$, in any dimension $d$. In ES's law
$T_0$ is given by the Coulomb energy of the localized state
$E_\xi=e^2/\xi$. The Mott and the ES laws, as well as the crossover
between the two regimes, were observed in various
experiments~\cite{Mott_ES_experiments}. Many qualitative features
that this complicated system exhibits can be obtained using rather
simple and heuristic arguments. However, using the single-particle
DOS in the calculation of the conductance, neglecting finite
temperature effects on the DOS and ignoring correlations between
energies and spatial locations is not discussed or can not be
justified in the framework of the heuristic arguments (see
[\onlinecite{Pollak_1_over_2}] for an elaborate discussion).

 In this work we study the crossover between Mott's and ES's laws,
using the theoretical framework of a Hartree
theory~\cite{coulomb_gap_mean_field,amir_glass}. The Hartree theory
takes into account interactions in a self-consistent manner. Hence,
it gives a well defined procedure for calculating various physical
properties and allows us to develop a clear intuitive physical
picture. Within the Hartree approximation the use of a renormalized
single-particle DOS is built in, with the electrons moving in an
effective potential due to the other electrons. While various other
theories neglect the interaction-induced correlations between
on-site energies and position \cite{{shegelski},{goedsche}}, these
exist in the theory we present: The mean-field equations
\cite{amir_glass} contain both the electronic site positions and the
renormalized energies, and therefore correlations between the two
exist in the self-consistent solution. We emphasize that within the
Hartree theory the average occupation and renormalized energy are
different at each electronic site. Albeit, the Hartree method is
uncontrolled as there is no small parameter in the theory. It is
therefore difficult to estimate rigorously its limit of validity.
Nevertheless, due to the long-range nature of the Coulomb
interactions, the potential energy at a given site is renormalized
by a large number of electronic sites. This suggests that the
approximation should be valid, and we believe it should be
applicable for a wide range of system parameters. Indeed, the
mean-field picture yields the well-established results for the
Coulomb gap in two dimensions \cite{amir_glass}, and captures well the results of aging experiments \cite{unpublished}  . 

In this work we demonstrate, in two dimensions, that the Hartree
theory yields the ES's and Mott's laws including the crossover
between them, for which we derive an analytical scaling relation
that holds under generic conditions, extending the results of Refs.
\cite{{ora_VRH},{meir_scaling},{aharony_sarachik},{rosenbaum}}. This
is a proof that interaction-induced correlations between location
and energy as well as finite temperature corrections are not
essential for determining the~conductance.

The Hartree approximation involves a numerical procedure.
Nevertheless, it clarifies our physical understanding of the complex
Coulomb glass. Its success in reproducing equilibrium (the gap in
the DOS) and near equilibrium properties (the conductance) and its
simplicity indicate that it may be usable in the estimates of
dynamical properties. Indeed, this was done in
[\onlinecite{unpublished}].

The structure of the paper is as follows. We first present the
model, and the heuristic arguments leading to the Mott and ES laws.
We show the Hartree theory gives a Coulomb gap consistent with other
theoretical and numerical approaches, and study the temperature
dependence of the DOS. We then explain the mean-field approach to
determining the conductivity, and the numerical procedure involved.
Finally, we analyze the crossover between the two regimes
analytically, under certain assumptions, and compare this to the
numerical results. The discussion that follows is centered around
the two-dimensional case, except for the scaling analysis which is
done for completeness in arbitrary~dimension.

 \section{The model of
hopping between localized states}

 The model analyzed consists of $N$
localized states and $M < N$ interacting electrons, with a coupling
between the electrons and a phonon reservoir \cite{amir_glass}. The
states are localized at sites ${r_i},\; i=1..N$, and by assumption
the average distance between the states is larger than the
localization length length $\xi$. The states have different on-site
energies, $\epsilon_i$, due to the disorder in the hosting lattice.
For simplicity, we will neglect in our analysis fluctuations in
$\xi$ from site to site. Since the states are localized, the
electrons will interact via an unscreened Coulomb potential $e^2/
|r_j-r_i|$ (to take into account the dielectric constant $\kappa$,
one should replace $e^2$ by $\frac{e^2}{\kappa}$ throughout). The
disorder bandwidth $W$, will be assumed to be much larger than the
Coulomb energy term $\frac{e^2}{r_{nn}}$ for the transport
calculations    , with $r_{nn}$ the average nearest-neighbor
distance. In the self-consistent Hartree approximation the energies
of the states are renormalized from $\epsilon_i$ to an energy $E_i$
due to the interaction with the other electrons.

\begin{figure}[b!]
\includegraphics[width=0.5\textwidth]{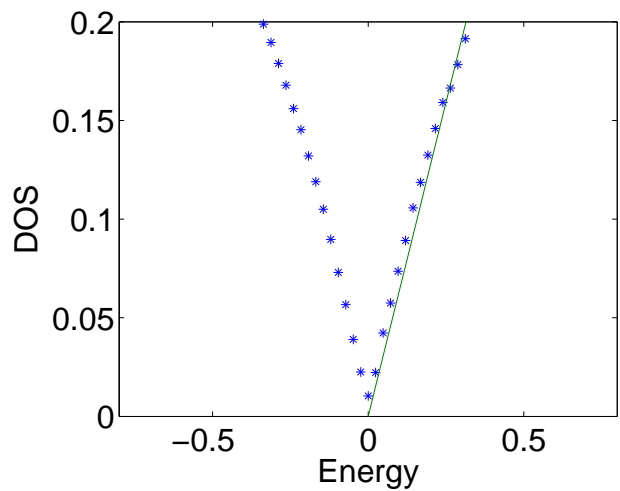}
\caption{A comparison between the Coulomb gap obtained by the
mean-field approximation, in two dimensions, and the theoretical
prediction \cite{efros_SC} of a linear gap with slope $\frac{2}{\pi
e^4}$. The number of sites used was N=10000, and the Fermi energy
was $E_f=0$ (half filling). The sites were uniformly distributed in
a square, with ${e^2}n^{\frac{1}{2}}=1$, where $n$ is the density.
The temperature was taken to be zero. The on-site disorder was
distributed uniformly in the interval $[-\frac{1}{2},\frac{1}{2}]$.
The y axis denotes the probability density of the energies $E_i$.
The points are an average over 300 instances.
}\label{coulomb_gap_2d_efros}
\end{figure}

Let us denote the energy difference of the electronic system before
and after a phonon-induced tunneling event by $E_{ij}=E_j-E_i$, and
the distance between their locations by $r_{ij}=|r_j-r_i|$. For weak
electron-phonon coupling the transition rate $\gamma^0_{ij}$ of an
electron from site $i$ to
site $j$, is approximately (for $E_{ij}<0$)\cite {efros}: %

\begin{equation}
\gamma^0_{ij} \sim |M_q|^2 \nu_p f\left(E_i\right)
\left(1-f\left(E_j\right)\right)
e^{-\frac{r_{ij}}{\xi}}[1+n(|E_{ij}|)], \label {rates}
\end{equation}
where $f(E)$ and $n(E)$ are the Fermi-Dirac and the Bose-Einstein
distributions, respectively; $M_q$ the electron-phonon coupling
constant and $\nu_p$ the phononic density-of-states. For upward
transitions ($E_{ij}>0$) the square brackets are replaced by
$n(|E_{ij}|)$ . We emphasize that the energy difference contains the
Coulomb interaction with all other charges. The $e^2/r$ interaction
term present in the Efros-Shklovskii argument should not be included
here, as in the master equation approach the charge is continuously
transferred. Indeed, only without this term will the approach lead
to detailed balance at equilibrium and to the correct result for the
Coulomb gap \cite{amir_glass}, including the coefficient
$\frac{2}{\pi e^4}$, as demonstrated in Fig.
\ref{coulomb_gap_2d_efros}. Recent numerical simulations have
confirmed the expected form of the three dimensional Coulomb gap
\cite {goethe}, and it would be interesting to see this also for the
two dimensional case.

\section{Analysis using heuristic arguments}

 The heuristic
arguments are based upon the low temperature approximation that the
conductance follows \cite{efros}:

\begin{equation}
\label{eq:conductance} \sigma \propto \gamma^0_{ij} \propto
e^{-\left(r_{ij}/\xi-|E_{ij}|/T\right)}.
\end{equation}

Where $\gamma^0_{ij}$ are two optimally coupled sites (i.e, with the
highest rate between them). To maximize exponential functions it is
sufficient to minimize the exponent even if the prefactors are not
optimized. Indeed, it pays sometimes for the electron to hop a
larger distance, thereby finding a state closer in energy and
reducing the necessary inelastic energy transfer \cite{mott}.

In the Hartree self consistent treatment it is legitimate to use the
non-interacting concept of DOS and to ask how interactions modify
it. To determine the shape of the interacting DOS, $\nu(E)$, (not to
be confused with the non-interacting DOS $\nu_0$) we assume that the
noninteracting and interacting energy-distance relation are equal,
i.e., for two dimensions $E=e^2/r=1/\nu(E) r^2$. This yields the ES
result  $\nu(e) \sim E/e^4$. The tunneling DOS is suppressed at low
energies due to interactions
\cite{{pollak}_,{efros2},{sarvestani},{vojta},{pankov}}. Notice that
$\nu(E) \sim \nu_0$ at $E_C = e^4 \nu_0$, an energy above which the
DOS becomes $\nu_0$. Hence $E_C$ determines the scale of the Coulomb
gap.

Regarding the conductivity of the system, a-priori there could be
four energy scales involved: the
 bandwidth $W$; the Coulomb energy $\frac{e^2}{r_{nn}}$, with $ r_{nn}$ the average
 nearest-neighbor distance; the Coulomb energy over the
 localization length distance $E_{\xi} \equiv \frac{e^2}{\xi}$;
 and $\delta_{\xi} \equiv \frac{1}{\nu_0 \xi^2}$, the level
 spacing in a box of size $\xi$. However, it is clear that the
 conductance as well as other physical quantities can not
 depend on $W$. Even in the limit of $W\longrightarrow \infty$,
 while keeping the (non-interacting) DOS at the Fermi level,
 $\nu_0$, constant, the expressions should be finite, since the
 energies involved in transport processes should occur near the
 Fermi energy $E_F$. Since $\frac{e^2}{r_{nn}}=\frac{E_{\xi} \sqrt{W}}{\sqrt{\delta_{\xi}}}$, this
 scale is also ruled out. Indeed, it is possible to express all
 the physical properties using $E_{\xi}$ and $\delta_{\xi}$.
Using similar arguments we can understand the formula for $E_C$:
Since the DOS is a static property of the system, determined by
interaction energies on scales much larger than the localization
length $\xi$, it cannot depend on $\xi$. The only energy scale that
we can construct from $E_\xi$ and $\delta_\xi$ that does not depend
on $\xi$ is $E_\xi^2/\delta_\xi=E_C$.

Optimizing Eq.~(\ref{eq:conductance}) with the DOS described above
yields Mott's law for temperatures above a certain temperature and
ES's law \cite{efros} for lower temperatures. The crossover
temperature in two dimensions is given by :\be
T_x=E_\xi^3/\delta_\xi^2 \label {T_x}. \ee Note that this is
consistent with the physical picture given previously, where it was
shown that all physical properties must depend on $E_{\xi}$ and
$\delta_\xi$.

So far we have reviewed simple heuristic arguments and discussed
Mott's and ES's laws in a non-rigorous way. This enabled us to
introduce the energy and length scale in the problem we study. In
the rest of the paper we will perform analytical and numerical
analysis of the Hartree approach, and show that it reproduces the
non-rigorous results although it includes temperature and
correlation effects that were omitted in the above heuristic
approach.

\section {Temperature dependence of the Coulomb gap}

 \label {t_dep}
Determining the value of the finite DOS at the Fermi energy due to
the effect of temperature is a problem which has not been settled
yet: numerical investigations by [\onlinecite{sarvestani}] seem to
be in disagreement with other theoretical predictions \cite{raikh}.
We hereby present the results of the mean-field theory.
We calculate the density of states (DOS) at the Fermi-energy in the
presence Coulomb interactions and finite temperature. For an
infinite system at zero-temperature, the Efros-Shklovskii Coulomb
gap will make the DOS vanish at the Fermi-energy. We know that
finite-size effects will give rise to a finite DOS, proportional to
$\frac{1}{\sqrt N}$\cite{coulomb_gap_mean_field}. Let us assume we
have a system large enough so that these finite-size effects will be
negligible compared to the effects of the finite temperature. The
finite-temperature mean-field equations are \cite {amir_glass}:

\be E_i = \epsilon_i + \sum_{j\neq i} \left(
\frac{1}{1+e^{\frac{E_j-E_F}{T}}}\ -Q_b \right) \frac{e^2}{r_{ij}}
 \label
{solve_E}. \ee For half filling, $E_F=0$, and the positive
background charge $Q_b=0.5$.

\begin{figure} \centering
\subfigure[] 
{
    \label{fig:sub:a}
    \includegraphics[width=0.4\textwidth]{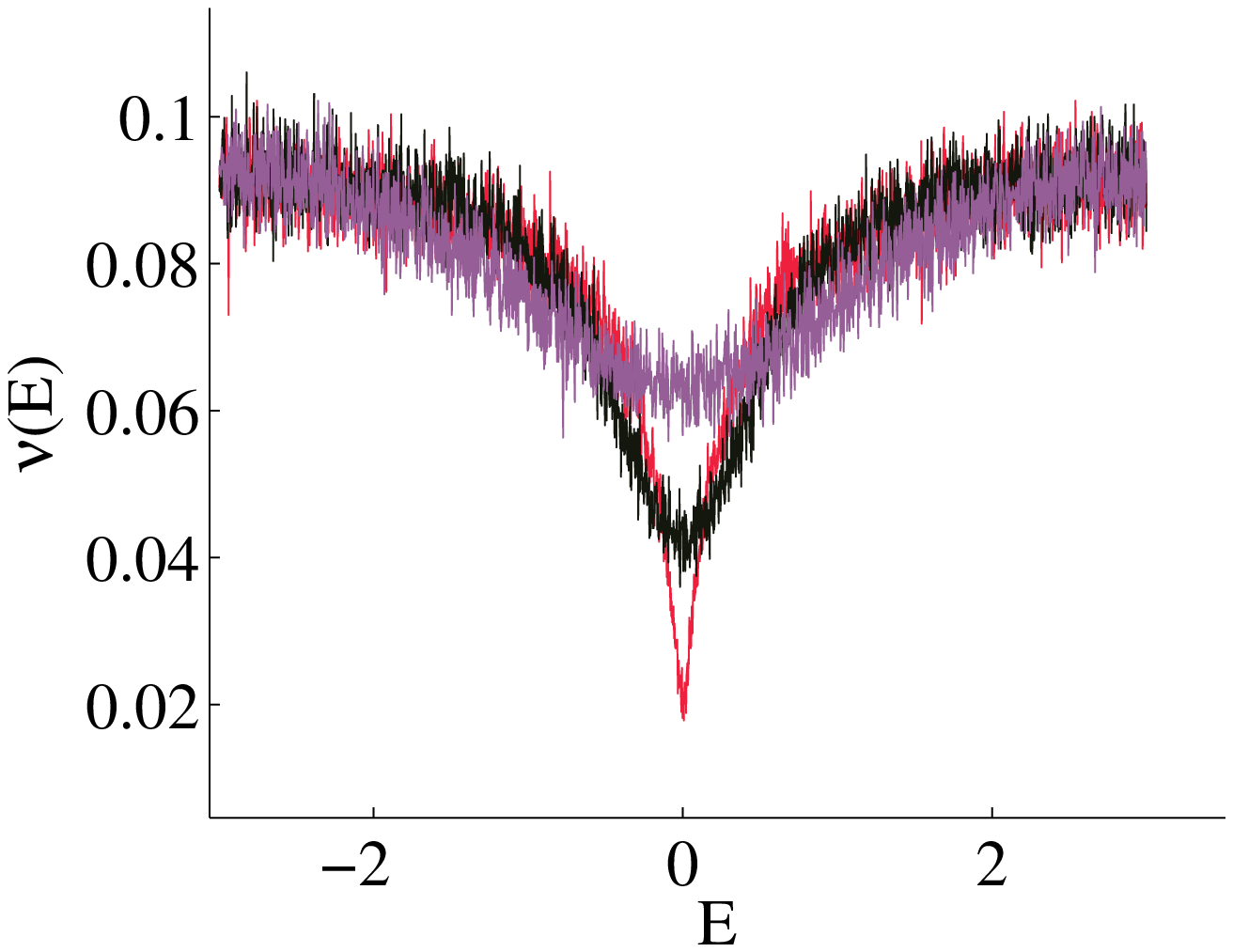}
} \hspace{1cm}
\subfigure[] 
{
    \label{fig:sub:b}
    \includegraphics[width=0.4\textwidth]{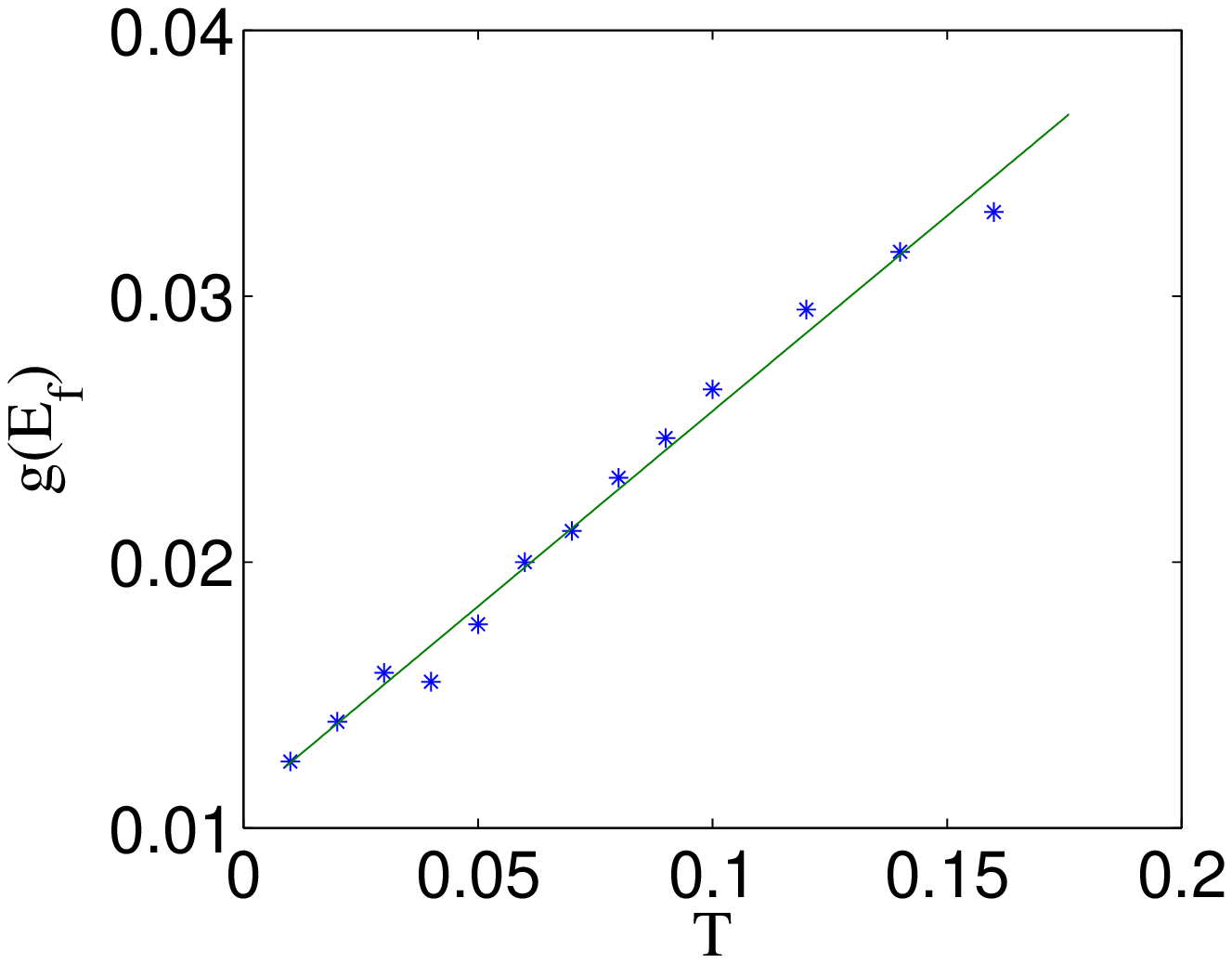}
} \caption{Density of states as a function of energy and
temperature. Each graph is an average over 2000 iterations. For
each, $N=1000$, and the ratio $\frac{T}{e^2 n^{\frac{1}{2}}}$, where
$T$ is the temperature, and $n$ the density, was varied between 0.01
and 0.58. The on-site disorder was taken uniform in the interval
$[-5,5]$.}
\label{T_dep} 
\end{figure}

We have solved these equations numerically in two dimensions, and
found a linear dependence of the DOS at $E_f$ as function of
temperature. This is consistent with ref. [\onlinecite{raikh}]. The
results are demonstrated in Fig. \ref{T_dep}. Thus $\nu(E_f)=\alpha
T$, and since the Coulomb gap is washed out at a temperature of
order $E_C$, we know that $\alpha \sim \frac{\nu_0}{E_C}$. For the
data corresponding to Fig. \ref{T_dep} we find that $\alpha \approx
0.15$. In this case the ratio $\frac{\nu_0}{E_C} \approx 0.1$, which
is consistent. We can now understand why this temperature dependence
does not affect the crossover discussed in the previous section:
while the finite-temperature contribution is linear in $T$, the
optimal hop energy scales as $\sqrt{T}$ at low temperatures (before
the crossover to Mott's regime). One can check that even at the
crossover temperature the finite-temperature correction to the DOS
is not important.

\section{Mean-field calculation of the conductance}

We calculated the
conductance numerically using the following scheme:

1. Find the equilibrium occupation number and energy associated with
each electronic site within the mean-field picture, using Eq.
(\ref{solve_E}). This takes the interactions into account, and the
density-of-states obtained will manifest the Coulomb gap
\cite{amir_glass}.

2. Construct the Miller-Abrahams resistor network
\cite{{miller_abrahams},{ambegaokar}}. For a non-interacting
problem, this is an \emph{exact} method to find the resistance of
the system by finding that of a certain resistor network. Here, the
resistor between the sites $i$ and $j$ is given by
$R_{ij}=\frac{kT}{e^2 \gamma^0_{ij}}$, where $\gamma^0_{ij}$ are the
equilibrium transition rates , which are calculated based on the
energies $E_i$ and occupation numbers of the sites, see
Eq.~(\ref{rates}). Two sites which are close to each other in space
and close in energy to the Fermi energy, will have a lower value of
resistance between them: if the sites are far away in space, the
overlap integrals will vanish exponentially. If the sites are far
from the Fermi energy, they will be permanently full or empty, and
will not contribute to the conductance.

3. Find the resistance of the system. This is done by solving for
the energies (voltages) at the sites. In steady-state, the sum of
currents into and out of each site must vanish. Expressing these
currents in terms of the resistor network values and the energies
yields a set of linear equations, solved numerically.

\begin{figure}[b!]
\includegraphics[width=0.5\textwidth]{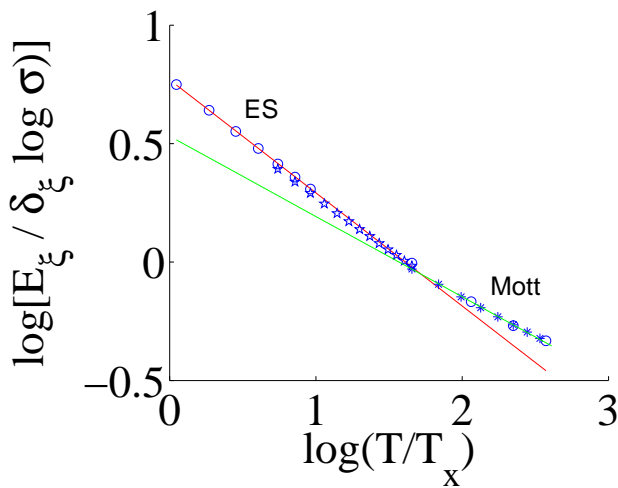}
\caption{ The crossover between ES and Mott VRH as the temperature
is varied. Using the scaling relation of Eq. (\ref{scaleit}), we
have collapsed data from different parameter sets onto the same
master curve. We use N between 500 and 3000, and each point
represents the average over 10 runs. We have verified that
logarithmic averaging yields essentially the same results. The sites
were chosen with uniform probability in a $\sqrt{N}X\sqrt{N}$ square
with density $n=1$, and the energy scale was chosen such that $e^2
n^{\frac{1}{2}}=1$. $n^{\frac{1}{2}} \xi $ was varied between 0.1
and 0.2, where $\xi$ is the localization length, and the on-site
disorder was distributed uniformly in the interval
$[-\frac{1}{2},\frac{1}{2}]$. The temperature axis was scaled
according to the crossover temperature of Eq. (\ref{T_x}). The
conductance was scaled according to the scaling relation of Eq.
(\ref{scaling}) The lower temperature fit has an exponent of $0.49
\pm 0.02$, consistent with the Efros-Shklovskii value of $1/2$ while
the higher temperature fit has an exponent of $0.34 \pm 0.01$,
consistent with the Mott value of $1/3$. The lines are the best fits
for the relevant data sets. In the Mott regime where the linear fit
was taken the conductance changes by more than 3 orders of
magnitude. In the ES regime, the conductance changes by more than 6
orders of magnitude. \label{crossover_fit} }
\end{figure}

Executing the numerical procedure, we found that above a certain
crossover temperature $T_x$, given by Eq. (\ref{T_x}), approximately
$e^{-{(\frac{T_0}{T})}^\zeta}$ behavior with $\zeta \sim 0.34 \pm
0.01$ was observed for the resistivity, close to the Mott value in
two dimensions of $\frac{1}{3}$. Below this temperature $\zeta \sim
0.49 \pm 0.02$ was observed, in accordance with the Efros-Shklvoskii
result \cite{finitesize}. In the next section we give a scaling
argument, which shows that: \be \rm{log} \sigma \sim
\delta_\xi/E_\xi \emph{f}( \emph{T}_\emph{x}/\emph{T} ).
\label{scaleit} \ee This allows us to collapse numerical data
obtained for different sets of parameters. The crossover between the
Mott and Efros-Shklovskii regimes is manifested in the asymptotics
of $f(x)$. Fig.~\ref{crossover_fit} demonstrates the crossover
between the two regimes, and the data collapse, validating the
scaling relation. We should emphasize, however, that the scaling
relation is derived under certain assumptions: we assume that the
temperature is low enough such that we do not have the trivial
scenario of nearest-neighbor hopping. At $T>W$, we shall have
nearest-neighbor hopping, and Mott's law will not be valid: this
gives a 'smoothing off' of the Mott exponent, to a temperature
independent regime. Thus, we do not expect complete data collapse
for the regime where the temperature is larger than the disorder.
Additionally, at very low temperatures, finite-size effects give
rise to a finite DOS, and thus give rise to deviations from the
Efros-Shklovskii hopping mechanism. Nevertheless,
Fig.~\ref{crossover_fit} shows clearly that data collapse is
obtained. The least-squares fits in the asymptotic (linear) regimes
gave the exponents discussed previously.

\section {Form of the crossover} \label {form_crossover} The
crossover between the Mott and ES regimes was discussed in
references
\cite{{ora_VRH},{meir_scaling},{aharony_sarachik},{rosenbaum}}.

We shall now calculate the form of the crossover within our
theoretical framework. We shall show that scaling is obeyed, for any
dimensions, regardless of the details of the Coulomb gap. This
generalizes previous work, for any DOS which depends only on the
parameters $e^2$ and $\nu_0$. As previously explained, this is the
generic form for the DOS. Notice that in principle the DOS is weakly
dependent on temperature, and therefore this scaling relation is
only an approximation. Let us proceed to the proof, for any
dimension $d$. The basic equation connecting the energy $E$ and
distance $r$ associated with a hop is obtained by requiring that the
average number of sites within an area of order $r^2$ and energy
lower than $E$ should be of order unity. Denoting the DOS as
$\nu(E)$, this gives:

 \be r^2 \int_0^E \nu(E')dE' \sim 1 .\label{E_from_r}\ee

 The second equation comes from optimizing $\eta=
\frac{r}{\xi}+\frac{E}{T} \label {eta} $. The optimal $\eta$ will
determine the conductance of the system via $\log \sigma \sim
-\eta$. Differentiating the first equation with respect to $r$ gives
\be \nu(E)= \frac{-d}{r^{d+1}} \frac{dr}{dE}. \label {eq1} \ee
Demanding the derivative of $\eta$ with respect to $r$ to vanish
gives \be \frac{1}{\xi}+ \frac{1}{T}\frac{dE}{dr} =0 .\label
{eq2}\ee Combining the last two equations yields

\be \nu(E)= \frac{\xi d}{\left[r\left(E\right)\right]^{d+1} T},
\label{eq4} \ee therefore: \be \int^E \nu(E') dE' =
\left(\frac{\nu(E) T}{d \xi}\right)^{\frac{d}{d+1}} . \label {eq3}
\ee To proceed, it is useful to define the function
$\psi(E)=\frac{\int^E \nu(E')
dE'}{({\frac{\nu(E)}{2})}^{\frac{d}{d+1}}}$. We have by assumption
$\psi= \psi(E, \nu_0, e^2)$, and we obtain the equation: \be
\psi(E,\nu_0,e^2)= {\left(\frac{T}{\xi}\right)}^{\frac{d}{d+1}}.
\label{L} \ee Since the dimensions of the LHS are
$[energy/length]^{\frac{d}{d+1}}$, $\psi$ must have the form: $
\psi= g(m) [E^{\frac{2}{d}} \nu_0^{\frac{2}{d}}
e^2]^{\frac{d}{d+1}},$ where $m=\frac{E}{ e^{\frac{2d}{d-1}}
\nu_0^{\frac{1}{d-1}}}$ is a dimensionless combination of the
parameters of $\psi$, and $g(m)$ is a certain function. Then the
equation takes the form:

\be \left(\frac{E}{ e^{\frac{2d}{d-1}}
\nu_0^{\frac{1}{d-1}}}\right)=g^{-1}\left[
{\left(\frac{T}{E^{\frac{2}{d}} \nu_0^{\frac{2}{d}} e^2
\xi}\right)}^{\frac{d}{d+1}} \right].\ee

We have an equation of the form $A E = f (B E),$ with $A=
\frac{1}{e^{\frac{2d}{d-1}} \nu_0^{\frac{1}{d-1}}}$ and $B= e^d
\nu_0 (\frac{\xi}{T})^{\frac{d}{2}}$. Both $A E$ and $B E$ are
dimensionless. The general solution will give us $E= \frac{1}{A}
\phi(\frac{B}{A})$, for a certain function $\phi$. Plugging in the
values of $A$ and $B$, we obtain \be \frac{E}{T}= \frac{1}{(e^2
\nu_0)^{\frac{1}{d-1}} \xi} \theta(\frac{T_x}{T}) , \label{E_T}\ee
with $T_x$ as defined by Eq. (\ref{T_x}), and $\theta$ is a
non-universal function, depending on the form of the DOS.

 A similar dimensional analysis shows that $\frac{r}{\xi}$ follows
 an identical scaling law (but with a different function).
 Using the definition of $\eta$, we reach the conclusion that the
conductance follows the proposed scaling, which can be written in
terms of the energy scales $\delta_{\xi}$ and $E_{\xi}$ as: \be \log
\sigma \sim \left( \frac{\delta_{\xi}}{E_{\xi}} \right)
f\left[\left(E_{\xi}^{d+1}/{\delta_{\xi}^2}\right)^{d-1}/{T}\right].
\label {scaling}\ee

Notice that for $d=2$ we retrieve the crossover temperature of Eq.
(\ref{T_x}). The analysis did not assume anything on the shape of
the DOS, other than the assumption that it is a function of $\nu_0$
and $e^2$: it does not have to contain a plateau or a power-law
Coulomb gap. Nevertheless, if it does contain these features, $f$
must have asymptotics which correspond to the Mott and ES VRH: at $x
\gg 1$ we have $f \propto x^{\frac{1}{d+1}}$, while for $x \ll 1$ we
have $f \propto x^{\frac{1}{2}}$. An advantage of this calculation
compared to other theories is that it allows one to complete the
calculation for \emph{any} DOS. Hence, it can be used by
experimentalists to 'reverse-engineer' and find the function $f$
that described the DOS from the conductivity measurements.

\section{Summary}

We have given a consistent Hartree-theory for the conductance of the
electron glass. The long-range nature of the interactions should
justify this approximation. In a previous work we showed the theory
gives a linear Coulomb gap at zero temperature \cite{amir_glass}, in
accordance with other theories. Since the interactions are treated
on a mean-field level, we can still use the single-particle DOS to
characterize the conductance. This explains why the Coulomb gap
indeed affects the conductance. Indeed, by numerically solving the
Miller-Abrahams resistor network for the system, we found a
crossover between the ES and Mott VRH regimes. Scaling is
generically obeyed at the crossover, which we have derived
analytically (neglecting interaction induced correlations), and
compared to the numerics (which takes them into account).  In future
work we intend to incorporate many-electron tunneling into the
theory, as well as study the out-of-equilibrium properties of the
system.

We thank A. Aharony, A.L. Efros, O. Entin-Wohlman, Z. Ovadyahu and
M. Pollak for useful discussions. This work was supported by a BMBF
DIP grant as well as by ISF and BSF grants and the Center of
Excellence Program. A.A. acknowledges funding by the Israel Ministry
of Science and Technology via the Eshkol scholarship program.


\begin{thebibliography}{10}

\bibitem{efros}
B. Shklovskii and A. Efros, {\em Electronic properties of doped
semiconductors}
  (Springer-Verlag, Berlin, 1984).

\bibitem{efrosbook2}
{\em Electron-electron interactions in disordered systems}, edited
by A.~L.
  Efros and M. Pollak (North-Holland, Amsterdam, 1985).

\bibitem{Gantmakherbook}
V.~F. Gantmakher, {\em Electrons and Disorder in Solids} (Oxford
University
  Press, Oxford, 2005).

\bibitem{pollakbook1}
{\em Hopping Transport in Solids}, edited by M. Pollak and B.~I.
Shklovskii
  (Elsevier, London, 1990).

\bibitem{mott}
N.~F. Mott, Phil. Mag. {\bf 19},  835  (1969).

\bibitem{pollak_}
M. Pollak, Discuss. Faraday Soc. {\bf 50},  13  (1970).

\bibitem{efros2}
A.~L. Efros and B.~I. Shklovskii, J. Phys. C {\bf 8},  L49  (1975).

\bibitem{coulomb_gap_experimental}
J.~G. Massey and M. Lee, Phys. Rev. Lett. {\bf 75},  4266  (1995).

\bibitem{butko}
V.~Y. {Butko}, J.~F. {DiTusa}, and P.~W. {Adams}, Physical Review
Letters {\bf
  84},  1543  (2000).

\bibitem{Mott_ES_experiments}
A. Mobius et al., J. Phys. C 16, 6491 (1983), A. Mobius et al., J.
Phys. C 18,
  3337 (1985), A. Mobius, J. Phys. C 18, 4639 (1985), W. N. Shafarman, D. W.
  Koon, and T. G. Castner, Phys. Rev. B 40, 1216 (1989), Y. Zhang, P. Dai, M.
  Levy, and M. P. Sarachik, Phys. Rev. Lett. 64, 2687 (1990), R. Rosenbaum,
  Phys. Rev. B. 44, 3599 (1991), Y. Liu, B. Nease, K. A. McGreer, and A. M.
  Goldman, Europhys. Lett. 19, 409 (1992) I. Shlimak, M. Kaveh, M. Yosefin, M. Lea and P. Fozooni, Phys. Rev. Lett.
  68, 3076 (1992), U. Kabasawa et al., Phys. Rev. Lett. 70, 1700 (1993), J.
  Lam, M. D'Iorio, D. Brown, and H. Lafontaine, Phys. Rev. B 56, R12741 (1997),
  M. P. Sarachik and P. Dai, Europhys. Lett. 59, 100 (2002).

\bibitem{Pollak_1_over_2}
M. Pollak, Phys. Stat. Sol. (c) {\bf 5},  667   (2008).

\bibitem{coulomb_gap_mean_field}
M. Grunewald, B. Pohlmann, L. Schweitzer, and D. Wurtz, J. Phys. C:
Solid State
  Phys., {\bf 15},  L1153  (1982).

\bibitem{amir_glass}
A. Amir, Y. Oreg, and Y. Imry, Phys. Rev. B {\bf 77},  165207
(2008).

\bibitem{shegelski}
M.~R.~A. Shegelski and D.~S. Zimmerman, Phys. Rev. B {\bf 39},
13411   (1989).

\bibitem{goedsche}
F. Goedsche, phys. stat. sol. (b) {\bf 140},  225  (1987).

\bibitem{unpublished}
A. Amir, Y. Oreg, and Y. Imry, to appear in PRL.

\bibitem{ora_VRH}
O. Entin-Wohlman, Y. Gefen, and Y. Shapira, J. Phys. C {\bf 16},
1161  (1983).

\bibitem{meir_scaling}
Y. Meir, Phys. Rev. Lett. {\bf 77},  5265   (1996).

\bibitem{aharony_sarachik}
A. Aharony, Y. Zhang, and M.~P. Sarachik, Phys. Rev. Lett. {\bf 68},
3900
  (1992).

\bibitem{rosenbaum}
R. Rosenbaum, N.~V. Lien, M.~R. Graham, and M. Witcomb, J. Phys.:
Condens.
  Matter {\bf 9},  6247  (1997).

\bibitem{goethe}
M. Goethe and M. Palassini, Phys. Rev. Lett. {\bf 103},  045702
(2009).

\bibitem{efros_SC}
A.~L. Efros, J. Phys. C: Solid State Phys {\bf 9},  2021  (1976).

\bibitem{sarvestani}
M. Sarvestani, M. Schreiber, and T. Vojta, Phys. Rev. B {\bf 52},
R3820
  (1995).

\bibitem{vojta}
F. Epperlein, M. Schreiber, and T. Vojta, Phys. Rev. B {\bf 56},
5890
  (1997).

\bibitem{pankov}
M. Muller and S. Pankov, Phys. Rev. B. {\bf 75},  144201  (2007).

\bibitem{raikh}
A.~A. Mogilyanski and M.~E. Raikh, Sov. Phys. JETP {\bf 68},  1081
(1989).

\bibitem{miller_abrahams}
A. Miller and E. Abrahams, Phys. Rev. {\bf 120},  745  (1960).

\bibitem{ambegaokar}
V. Ambegaokar, B.~I. Halperin, and J.~S. Langer, Phys. Rev. B {\bf
4},  2612
  (1971).

\bibitem{finitesize}
At low temperatures finite size effects will give rise to a finite
DOS at the
  Fermi energy, modifying this picture and raising the conductance. These were
  also seen numerically.

\end{thebibliography}
\end{document}